\documentclass[journal]{IEEEtran}

\usepackage{graphicx}
\usepackage{epstopdf}  
\usepackage{svg} 
\usepackage{cite}
\usepackage{amsmath, amssymb, mathtools}
\usepackage{mathrsfs}
\usepackage{url}
\usepackage{tikz}
\usetikzlibrary{positioning, arrows.meta, calc}
\usepackage{float}
\usepackage[table]{xcolor}
\usepackage{array}
\usepackage{booktabs}
\usepackage{graphicx}  
\usepackage{subcaption}
\usepackage{booktabs}
\usepackage{multirow}
\usepackage{siunitx}
\usepackage{cleveref} 
\usepackage{pgfplots}
\pgfplotsset{compat=1.18}
\usepackage{tikz}
\definecolor{headerblue}{RGB}{93,190,230}
\definecolor{rowblue}{RGB}{220,240,250}
\definecolor{lightblue}{RGB}{235,247,253}

\begin{document}
\title{Transient Stability of Offshore Energy Hubs}
\author{
Alban J.~F.~Duvivier,
Dominic Gro\ss{},~\IEEEmembership{Senior Member,~IEEE},
Daniel M\"uller,~\IEEEmembership{Member,~IEEE}
and Nicolaos A.~Cutululis,~\IEEEmembership{Senior Member,~IEEE}%
\thanks{
This work is part of the Offshore Energy Hub research project funded by EUDP
(Energy Technology Development and Demonstration Program), case no. 64022-1011.
}%
\thanks{
A.~J.~F.~Duvivier, D.~M\"uller, and N.~A.~Cutululis are with the
Department of Wind and Energy Systems, Technical University of Denmark (DTU), Denmark.
}%
\thanks{
D.~Gro\ss{} is with the Department of Electrical and Computer Engineering,
University of Wisconsin–Madison, Madison, WI 53706, USA
(e-mail: dominic.gross@wisc.edu).
}
}

\maketitle
\begin{abstract}
Offshore energy hubs (OEHs) use grid-forming modular multilevel converters (MMCs) to enable large-scale offshore wind integration and multi-terminal HVDC operation. In HVDC-connected offshore wind farms and OEHs, the offshore grid-forming HVDC converters absorb active power from an offshore AC grid supplied by the wind farms and convert it to DC power for transmission to the onshore grid. Converter current limiting under different fault types in this setting is an understudied topic in the literature, which mostly focuses on power-injecting converters.
This paper proposes a unified current-limiting strategy that combines a variable virtual impedance (VVI), based on a smooth threshold function, with a novel virtual-power (VP) mechanism derived from the power dissipated in the virtual resistance. The VVI ensures current limitation during fault-induced overcurrents while preserving voltage-source behavior, whereas the VP mechanism adds a compensating power term into the synchronization loop, enabling automatic power redistribution among converters.
$P-\delta$ analysis further shows that a more resistive VVI can improve the transient stability of power-absorbing converters, while the proposed VP mechanism further enlarges the stability margin. EMT simulations validate that the combined VVI–VP strategy limits fault currents, maintains synchronism during severe faults, and achieves coordinated post-fault power sharing in fully converter-based OEHs.
\end{abstract}


\subsection*{Nomenclature}
\begin{description}
\item[VVI] Variable Virtual Impedance
\item[VP] Virtual Power
\item[EMT] Electromagnetic Transient
\item[MMC] Modular Multilevel Converter
\item[OEH] Offshore Energy Hub
\end{description}


\section{Introduction}
\IEEEPARstart{T}{he} rapid increase in renewable-based electrical power generation, driven by European energy agreements and carbon-neutrality targets~\cite{Directorate-GeneralforEnergy2024MemberEnergy}, has accelerated the transition toward power systems dominated by power electronic converters. As conventional synchronous generation is progressively displaced, new technologies are required to transport and integrate decentralized power sources while maintaining system stability.
A prominent example of this evolution is the development of offshore energy hubs (OEH)~\cite{Energinet2023OffshoreRe-quirements}, which aggregate offshore renewable generation and interconnect it with onshore grids via converter-based AC/DC infrastructure.

\begin{figure}
    \centering
    \includegraphics[width=\linewidth]{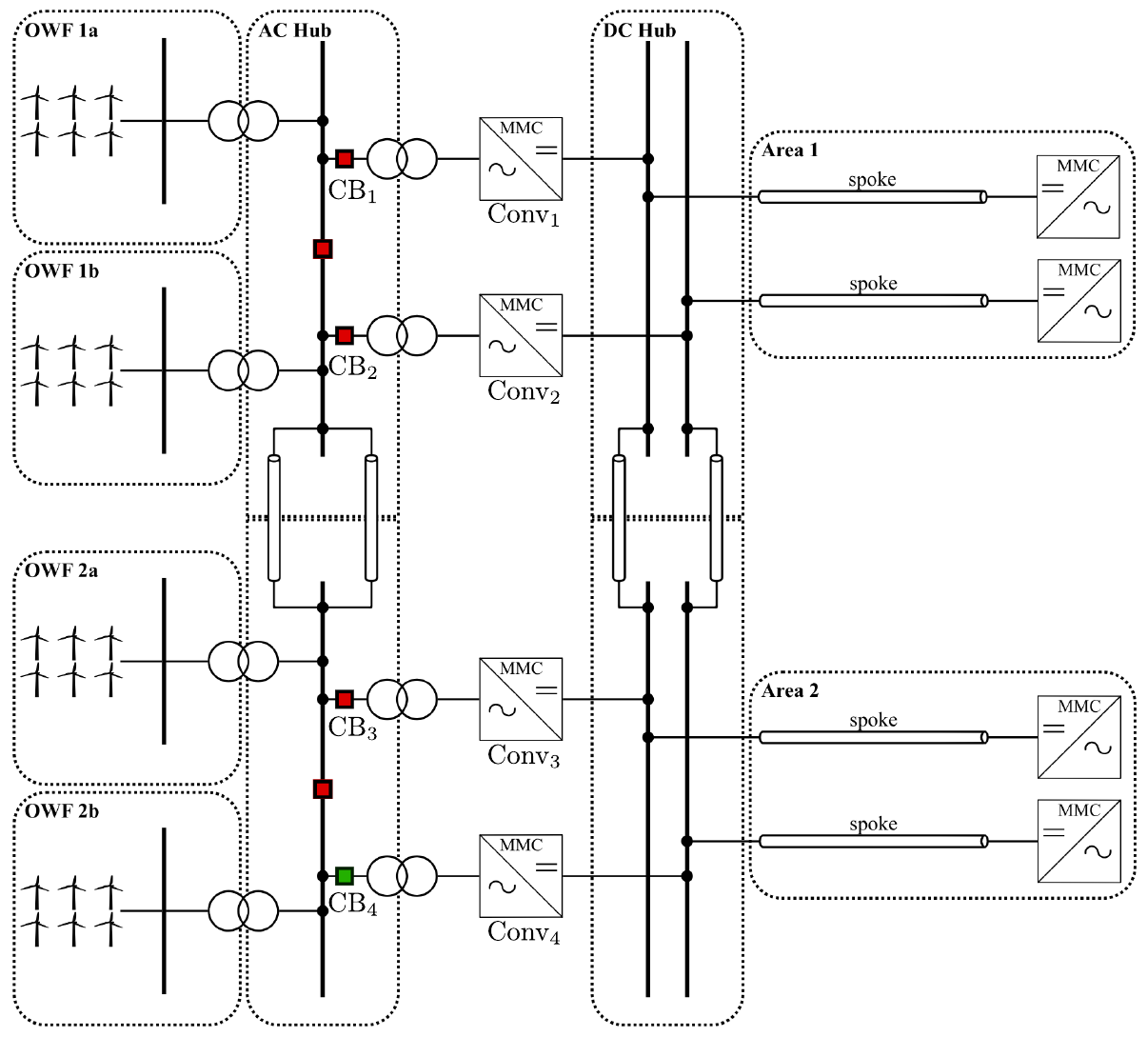}
    \caption{Hybrid topology for OEH connecting 4x $1$~GW wind farms to 2 onshore systems through 2 HVDC bi-pole connections~\cite{mueller2024oeh}. The green breaker (CB$_4$) indicates the disconnected pole for the study case in Sec.~\ref{sec:CaseStudy}.}            
    \label{fig:OEH}
\end{figure}

A schematic of an OEH can be found in Fig.~\ref{fig:OEH}. It includes four aggregated $1$-GW wind farms and four $1$-GW HVDC converters, split into two bi-poles. Among the conceptual OEH configurations, the hybrid hub allows components to be interconnected on either the AC or DC side. The study carried out in this paper focuses on the AC-connected hub, allowing current to be split between the converters on the AC side during converter failure.

From a system perspective, OEHs represent emerging converter-dominated power systems. In this environment, the absence of large synchronous machines implies that inertial response and frequency/voltage regulation are no longer inherently available and must instead be provided by grid-forming converters~\cite{Misyris2018NorthAssessment, mueller2024oeh}.

A key challenge in such systems is the fault response of the power electronic converters that form the OEH grid. Unlike synchronous machines, which can supply high short-circuit currents, power converters are inherently current-limited due to the thermal limits of their semiconductor switches. This constraint becomes critical during grid faults, directly affecting transient stability, synchronization, and fault ride-through capability. Effective current-limiting strategies are therefore required to preserve grid-forming behavior, ensure harmonic stability during fault inception, and re-synchronization after fault clearance.

Various current-limiting methods have recently been studied in numerous papers.
As explained in~\cite{Baeckeland2024OvercurrentDiscussion, Ordono2024CurrentThrough, Taul2020CurrentConditions, Qoria2020CurrentVSCs}, there is general agreement that current limiters for grid-forming controls can be categorized into direct and indirect limiters. 
Direct current limiters operate within an inner current loop, typically by saturating the current reference~\cite{Fan2022ADisturbances, Huang2019TransientLimitation}. They exploit the high bandwidth of the inner current controller to clamp the converter current within a few switching periods, yielding very accurate and fast limitation largely independent of the fault location. Once engaged, however, the limiter overrides the outer voltage, and grid-forming power/angle controllers and the voltage reference provided by the outer grid‑forming control are no longer fully tracked at the converter terminal. As a result, the converter temporarily loses its self-synchronizing voltage‑source behavior~\cite{Baeckeland2024OvercurrentDiscussion, Taul2020CurrentConditions}.

Another key disadvantage of direct limiters is that they require a cascaded control architecture that depends on adequate bandwidth separation between loops, which can be difficult to achieve in high-power applications operating at low switching frequencies~\cite{Ordono2024CurrentThrough}. 
Moreover, the tuning of the cascaded current/voltage controls has been shown to be highly dependent on grid strengths~\cite{Qoriatuning2018,grosstimescales2021} and significantly affects the system stability margin~\cite{Anbuselvi2016ImpactControl}.

Indirect current limiters, on the other hand, modulate variables such as AC voltage references (i.e., through virtual impedance), power set‑points, or voltage magnitude/frequency references rather than directly clamping the converter current references. This inevitably slows their response, because the limiter action must propagate through any inner current/voltage control loops. However, several studies show that such methods better preserve the grid‑forming, voltage‑source behavior and improve transient stability compared to purely direct reference current clamping~\cite{Baeckeland2024OvercurrentDiscussion, Qoria2020CriticalAlgorithm}.

Beyond direct and indirect methods, hybrid current limiting controls combine direct and indirect current limiters~\cite{Qoria2020CurrentVSCs}. The fast response of the direct limiter ensures that the current is constrained as soon as the fault occurs, whereas the VI contributes to improved transient stability afterward. In~\cite{Shen2023TransientStrategies}, a mode-switching model, which relies on a PLL during the fault to re-synchronize, is compared with the virtual impedance and the current reference limiter. However, its dependence on a PLL makes it unreliable for weak grids. Finally,~\cite{Huang2019TransientLimitation} adds the q-axis voltage component to the droop, hereby introducing an additional stabilizing term that improves the converter’s transient stability under current‑limiting conditions. Crucially, all the aforementioned works focus on power converters injecting power into an AC grid. Moreover, VVI exhibits performance limitations under severe over- and under-frequency conditions~\cite{grossCGFM2026}.

Building on these approaches, this paper (i) adopts a variable virtual impedance (VVI) as the primary current-limiting mechanism, and (ii) extends the $P-\delta$ analysis to converters that absorb power from an AC grid. In contrast to existing smooth thresholding implementations, the proposed VVI is based on a continuous barrier-type function that grows unbounded as the converter current approaches its maximum allowable value. This ensures a smooth transition into current-limiting operation while enforcing a strict current constraint, thereby preserving analytical tractability and avoiding discontinuities in the control action.

To overcome the limitations of VVI under severe overpower conditions, this paper further combines the proposed current-limiting mechanism with virtual power (VP). VP generally refers to a modified or synthetic power signal fed into the outer synchronization loop (e.g., droop, VSG, or power-synchronization control) of grid-forming converters, rather than the raw measured terminal power. For instance,~\cite{Ordono2024CurrentThrough, Desai2025ComparisonFaults, Kkuni2024EffectsSolution} use (synthetic) unconstrained current references in the synchronization loop together with direct current limitation. VP increases the synchronization power, thereby improving transient stability. 
Another method, developed in~\cite{Baeckeland2025TransientLimiting}, uses an additional term, referred to as fictitious power, defined as the difference between the unconstrained and constrained active output power.

Besides its role in transient stability, the proposed VP also functions as an autonomous power dispatch mechanism among converters, e.g., during severe overpower conditions. Power-sharing strategies have been widely studied in converter-based systems, as reviewed in~\cite{Han2016ReviewMicrogrids}. Broadly, two main approaches can be identified in the literature. The first approach adapts the converter droop gains to modify power output, as in~\cite{Ziaeinejad2020PowerResources}. However, altering droop coefficients for post-fault redistribution is undesirable, since droop coefficients are selected for power sharing during normal operation and also directly influence small-signal stability and dynamic performance~\cite{Raza2018Small-SignalSystems, Wang2018Small-SignalMethod}. 
A second approach regulates power sharing by shifting droop intercepts or adjusting power setpoints, either through coordinated control layers or adaptive mechanisms, as proposed in~\cite{Ganguly2023DroopSources, Wang2024DispatchingMode}. In~\cite{HoaThiPham2020PowerControl}, the line impedance is explicitly incorporated into the power computation to improve proportional load sharing accuracy.
In addition, virtual‑impedance‑based methods have been used to shape converter output impedance for proportional sharing. In~\cite{Ruiming2018PowerControl}, the virtual resistance is chosen much larger than the physical line impedance, effectively making the system behave as a resistive network so that power can be shared in proportion to the inverse of these resistances. However, using such large virtual resistances reduces voltage support and limits power‑transfer capability, making this approach less attractive for high‑performance operation.

In contrast, this paper introduces a novel virtual-power concept based on the power dissipated in the virtual resistance. Rather than modifying droop parameters or external setpoints or explicitly compensating line impedances, the proposed approach reshapes the power used in the synchronization loop by leveraging the power dissipated in the virtual impedance itself. Besides enhancing transient stability, it inherently enables autonomous power redistribution among converters, making it particularly suitable for converter-dominated systems.

OEHs are converter-dominated systems in which offshore AC networks collect wind generation and export power to onshore grids via HVDC links. In this configuration, grid-forming converters absorb power from the AC grid and inject it into the DC network. Depending on the disturbance, converters may face overcurrent due to faults or due to overpower resulting from power imbalances following faults and/or converter and line outages. Although various studies have addressed both topics, few have focused specifically on converter-based networks operating under power-absorbing conditions. As discussed in Sec.~\ref{sec:VVI}, this requires considering the negative region of the $P\text{-}\delta$ characteristic, leading to fundamentally different stability properties. Moreover, to the best of the authors’ knowledge, no single control strategy has been proposed that simultaneously addresses both fault-induced overcurrent and overpower conditions.

In summary, the main contributions of this paper are
\begin{itemize}
    \item A smooth variable virtual impedance design for current limitation without discontinuities. Unlike existing smooth thresholding approaches~\cite{Baeckeland2022Stationary-FrameLimiting}, the proposed formulation grows unbounded as the current approaches the limit, ensuring strict current enforcement.
    \item A novel virtual power concept, derived from VVI power dissipation, that enhances transient stability and enables coordinated post‑fault power sharing among converters without modifying droop gains.
    \item A comprehensive transient stability analysis, including $P-\delta$ curve interpretation for power‑absorbing converters, a regime rarely examined in existing literature.
    \item Demonstration through EMT simulations that the proposed VVI–VP framework preserves synchronism during severe faults and ensures autonomous, proportional power redistribution once the disturbance is cleared.
\end{itemize}

Together, these contributions provide a control strategy tailored to fully converter‑based grids, addressing both the immediate fault‑response requirements and the subsequent restoration of stable power flows.


\section{Problem Statement and Requirements}

\subsection{Offshore Energy Hubs}

An OEH represents a dense electricity generation cluster in which power is generated close to the consumer (the HVDC converter). Because it is a closed network, the effects of a grid event cannot spread across a large, diverse power grid, as they would in an onshore system. Instead, any power imbalance must be absorbed entirely by the limited number of converters within the hub, making the production-consumption balance inherently more fragile. The loss of a single converter or transformer forces the remaining units to redistribute a share of the total power, with no surrounding network to provide relief, motivating the need for dedicated current-limiting strategies for overpower conditions.

Furthermore, the compact geographical footprint of an OEH implies that the electrical lines are short and hence present low impedance, so a fault will result in very high fault currents at the converter terminals, highlighting the need for current limitation in the converter control.

\subsection{Fault response}
Faults in an electrical network can manifest in different ways and have different repercussions. One type involves a short circuit in the cables, resulting in a decrease in voltage on the AC network. During this period, the converters must remain connected and supply fault current in accordance with the grid code specifications. If the fault is temporary, a return to steady state is expected with minimal transients. However, if the fault persists, it may, in some cases, lead to a change in topology, requiring autonomous redistribution of converter powers to avoid overloading any one of them. 
In this complete case, the default can be divided into three phases.
\subsubsection{\textbf{Immediate Fault Response – Current Limitation}}
Immediately after a fault occurs, converters must enter current-limiting mode to protect the semiconductor switches. This transition must occur on a sub-cycle timescale: for example, the fault-mode controller in~\cite{Taul2020CurrentConditions} is activated within $1$~ms after fault detection. Fast-acting limitation mechanisms are therefore required to prevent sustained overcurrent while maintaining controllability of the voltage source.

\subsubsection{\textbf{Resynchronization Phase}}
During the disturbance, the angle and frequency dynamics of the converters become critical. Under severe voltage dips, the power–angle relationship becomes ill-defined, and strict synchronization with the grid voltage cannot be maintained. Instead, the objective is to ensure that the converter angles and frequencies do not drift excessively relative to each other. Loss of synchronism occurs when the angle, once the fault is cleared, exceeds the stability boundary, typically characterized by a full power inversion before returning to equilibrium.

\subsubsection{\textbf{Post-Fault Power Redistribution}}

Following fault clearance, some converters may be overloaded depending on their initial power setpoint, reference power, or distance from the disconnected converter.
This is particularly relevant in OEHs, where the electrical network is relatively sparse and has few alternative grid-connected elements to absorb or redistribute power, limiting the system’s ability to naturally handle power imbalances following a disturbance.
An effective control strategy must therefore ensure that excess power is automatically and proportionally shared among the healthy converters according to their available margins, while preserving synchronization and respecting current constraints. This situation is referred to as an overpower scenario, in which overcurrent is reached without changing the voltage level.


\section{Proposed Control Strategy}

To address the challenges identified above, this paper proposes a control strategy that combines variable VVI with a novel VP mechanism. The key idea is to inject the power dissipated in the virtual impedance into the synchronization loop, thereby reshaping the effective power-angle characteristic during disturbances. This approach enables fast current limitation while simultaneously improving transient stability and facilitating coordinated post-fault power redistribution among the remaining converters. The following subsections detail the control structure, the derivation of the VVI and the VP terms, and its dynamics during faults.

\subsection{Smooth Variable Virtual Impedance}

\begin{figure}
    \centering
    \includegraphics[width=1\linewidth]{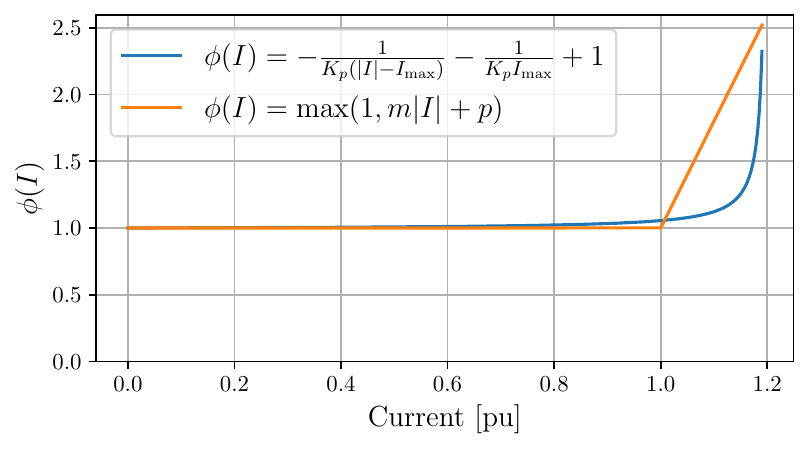}
    \caption{Comparison of the common VVI threshold function (orange) and proposed smooth barrier-type VVI function (blue). Notably, the barrier-type VVI function is smooth and $\phi(I) \to \infty$ as $I \to I_{\max}$ prevents the current from exceeding the limit.}
    \label{fig:thresholdFunction}
\end{figure}

A common implementation of VVI is presented in~\cite{Baeckeland2024OvercurrentDiscussion}. To this end, let $I$ denote the magnitude of the converter current. The initial virtual impedance is multiplied by a threshold function $\psi(I)$, which remains constant until the magnitude of the current $|I|$ reaches the threshold value $I_{\text{th}}$. It then increases linearly until $|I| = I_{\max}$, where $I_{\max} \in \mathbb{R}_{>0}$ is the maximum current and the value of $\psi(I_{\text{max}})$ is selected such that the current stays below the maximum allowable value during the worst-case scenario (a three-phase-to-ground fault at the converter terminals). This typical approach is shown in orange in Fig.~\ref{fig:thresholdFunction}. Previous work has also highlighted that using a non-linear $\psi$ enables a smooth transition between activating and deactivating the virtual impedance, thereby eliminating undesired transient phenomena~\cite{Baeckeland2024OvercurrentDiscussion, Gkountaras2015EvaluationMicrogrids}. Building on these insights, this paper uses smooth functions, denoted by $\phi$, to ensure smooth transitions between normal and current-limiting operation.

To enforce the current constraint, the non-linear barrier-type function 
\begin{align}
\label{eq:phi}
\phi(I) = -\frac{1}{K_{\text{p}} (|I| - I_{\text{max}})} - \frac{1}{K_{\text{p}} I_{\max}} + 1
\end{align}
is introduced, with $K_{\text{p}} \in \mathbb{R}_{>0}$. This function, illustrated in blue in Fig.~\ref{fig:thresholdFunction}, is designed such that $\phi(I)=1$ for $I=0$, thereby preserving the nominal virtual impedance under normal operating conditions. As the current magnitude increases, the function grows non-linearly following a hyperbolic profile. An asymptote is placed at $I=I_{\max}$, ensuring that the gain increases sharply as the current approaches its maximum allowable value. This results in a smooth but increasingly restrictive action, preventing the current from exceeding $I_{\max}$ while avoiding non-smooth control.

Finally, a factor $\sigma$ is added to the inductance term that corresponds to the ratio of the increase in the virtual inductance compared to the increase in the virtual resistance. The final equations of the VVI are given by
\begin{subequations}\label{eq:RXvI}
\begin{align}
R_{\text{VI}}(I) &\coloneqq \phi(I) R_{\text{VI},\text{init}}, \label{Requation} \\
X_{\text{VI}}(I) &\coloneqq \sigma \Big(\phi(I) - 1 + \frac{1}{\sigma}\Big) X_{\text{VI},\text{init}}, \label{Xequation}
\end{align}
\end{subequations}
where \eqref{Xequation} can be factorized as
\begin{align}
    X_{\text{VI}} = \sigma (\phi(I)-1) X_{\text{VI},\text{init}} + X_{\text{VI},\text{init}}
\end{align}
to better highlight the meaning of $\sigma$. The resulting VVI voltage drop is
\begin{align}
V_{\text{VVI},d,q} = R_{\text{VI}}(I) I_{d,q} \mp X_{\text{VI}}(I) I_{q,d}
\end{align}
which emulates the voltage drop produced by a series impedance connected to the grid.

Finally, a voltage controller 
\begin{subequations}\label{eq:VoltageController}
\begin{align}
E &= K_{\text{p}}(V_{\text{mea,flt}}-V_{\text{ref}}) + K_{\text{i}}(X_{\text{int}}), \label{equ:E}\\
\frac{d}{dt}X_{\text{int}} &= V_{\text{mea,flt}}-V_{\text{ref}},\\
V_{\text{ref}} &=  K_{q} (Q_{\text{set}} - Q_{\text{ref}}) + V_{\text{set}}.
\end{align}
\end{subequations}
provides the d-axis voltage reference for the modulation $E$, using a PI controller, while a $Q$-$V$ droop control gives the voltage reference $V_{\text{ref}}$ at the PCC, used as input to the PI controller.

\subsection{Virtual power using VVI power dissipation}
VVI has demonstrated its effectiveness in limiting converter current during fault conditions that result in low terminal voltages (i.e., short circuit faults). However, VVI may be less effective during overpower scenarios~\cite{grossCGFM2026}. Notably, once the fault is cleared and the faulted element is disconnected, an overpower situation is likely to occur in OEHs. Grid-forming converters electrically close to the faulted unit may become saturated as they aim to absorb the power previously exported by a disconnected converter (e.g., due to disconnecting a faulted line feeding the converter).

In this post-fault scenario, conventional VVI control is no longer suitable and may result in large voltage magnitude deviations and lack of well-defined power sharing. In traditional droop-controlled converter systems, after synchronization is re-established, power transfer among the $n$ remaining converters is governed exclusively by the droop equations
\begin{align}
\label{eq:omega}
\omega_i= \omega_0 + m_i (P_{i, \text{ref}} - P_{i,\text{meas}})
\end{align}
for all converters $i \in \{1,\ldots,n\}$, where $m_i \in \mathbb{R}_{>0}$ is the individual power to frequency droop value. The reference power and the droop value influence the power sharing among the converters in a proportional way, without constraining the share when reaching maximal values.

Under nominal steady-state conditions, all angular frequencies in \eqref{eq:omega} are synchronous (i.e., equal). Consequently, power sharing among converters is determined solely by the droop coefficients and the reference powers, i.e.,
\begin{align}
P_i = P_{i,\mathrm{ref}}
- \frac{1}{m_i}
\frac{
\displaystyle \sum\nolimits_{k=1}^N P_{k,\mathrm{ref}} - P_{\mathrm{WF}}
}{
\displaystyle \sum\nolimits_{k=1}^N \frac{1}{m_k}
},
\end{align}
for all $i \in \{1,\ldots,n\}$, where $P_{\text{WF}}$ is the total power production from the wind farms in the OEH.
During an overpower scenario, increasing the internal VVI does not modify the power sharing and therefore cannot reduce the current. Consequently, the VVI continues to increase with the current, ultimately leading to system collapse as it tends to infinity.

To overcome this limitation, the VVI's power dissipation is incorporated into the droop equation via the concept of virtual power. Since the converter current is predominantly aligned with the $d$-axis, only the resistive component of the virtual impedance contributes to active power dissipation. The associated power term can thus be expressed as $R_{\text{VI}} |I|^2$. This term virtually increases the power used in the synchronization loop, depending on the converter's loading and capacity, pushing out excess power to other converters. In order to cancel the effect during steady state operation, the $\phi$ function is lowered by one, removing the virtual losses. The power used in the synchronization is therefore defined as 
\begin{align}
    \label{VirtualPower}
    P_{\text{VP}} = P_{\text{mea}} - (\phi-1)R_{\text{VI,init}}  |I|^2
\end{align}
where $\phi(I)$ is the barrier function defined in \eqref{eq:phi} and $(\phi-1)R_{\text{VI,init}}  |I|^2$ is the (virtual) power dissipation of the VVI.

\begin{figure}[t]
    \centering
    \includegraphics[width=1\linewidth]{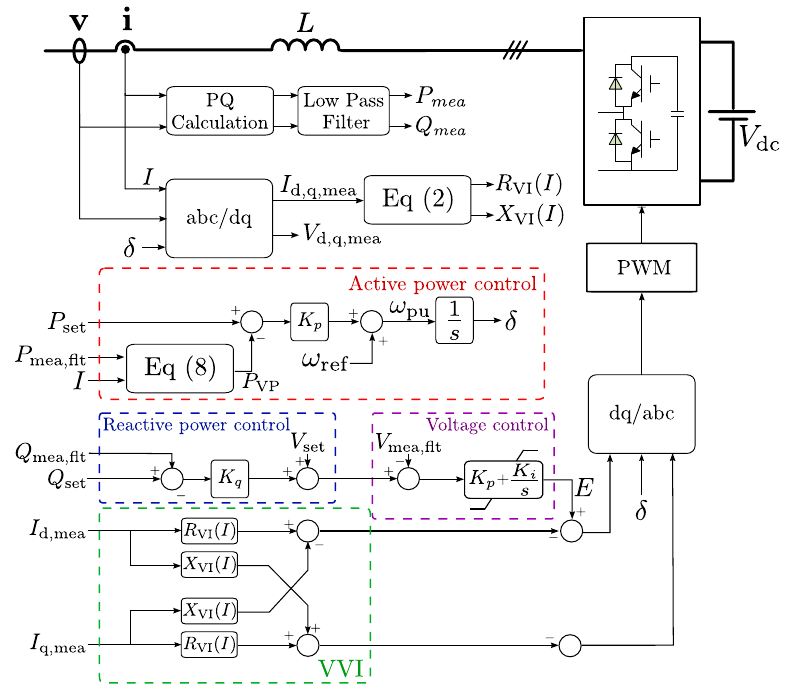}
    \caption{Control block diagram of the grid-forming converter comprising active power (red), reactive power (blue), voltage (purple), and Virtual Voltage Impedance (VVI, green) control loops. The active power loop uses droop control for grid synchronization via the angle $\delta$, while (2) computes current-dependent virtual impedance terms $R_\text{VI}(I)$ and $X_\text{VI}(I)$ to provide smooth current limiting and virtual power term.}
    \label{fig:blckdgrm}
\end{figure}

The overall control block diagram, including the VVI and the VP, is presented in Fig.~\ref{fig:blckdgrm}. The control has been implemented in PSCAD for the simulation studies in the subsequent sections.

\section{Stability and Power Transfer Analysis}

We first focus on transient stability and power transfer analysis of the proposed controller. Our analysis follows the fault chronology, starting with the $P-\delta$ curves and then proceeding to power dispatch after fault clearance.

\subsection{$P-\delta$ curve using VVI and VP}
$P-\delta$ curves are widely used in power system analysis to describe the relationship between transmitted active power and the power angle between voltage sources. They provide insight into synchronization stability limits in conventional generator-based systems. Although this work focuses on converter-based systems rather than classical voltage-source systems, the $P-\delta$ framework remains useful for building intuition and for better understanding the underlying power transfer and stability mechanisms~\cite{Huang2019TransientLimitation, Baeckeland2024OvercurrentDiscussion, Qoria2020CriticalAlgorithm, Ordono2024CurrentThrough, Desai2025ComparisonFaults, Kkuni2024EffectsSolution}.

\subsubsection{\textbf{Variable Virtual Impedance}}
\label{sec:VVI}

A key advantage of VVI is that the virtual impedance can be represented as an additional impedance at the converter output. The voltage reference generated by the $Q$–$V$ droop controller, $E$, then serves as the reference of the virtual voltage source behind that impedance. As the $Q$–$V$ droop variation is typically small, the voltage can be assumed constant during transients. Moreover, the PI controller in the voltage control is subject to saturation, so if the voltage reference increases excessively, the controller output $E$ in~\eqref{equ:E} remains clamped at its maximum value. The condition of a constant voltage source is required to derive the $P$–$\delta$ curves. However, it is generally not satisfied by direct current limiters or other limiting strategies, for which the assumption of a constant-voltage source is invalid.

\begin{figure}[htbp]
    \centering
    \includegraphics[width=1\linewidth]{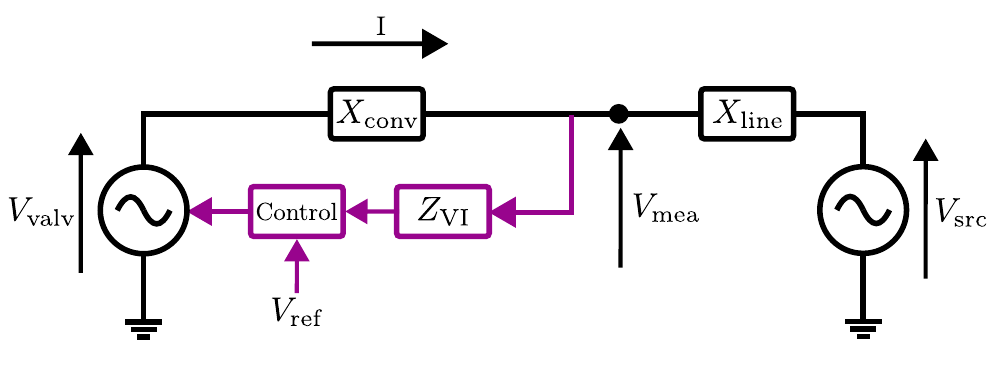}
    \caption{Circuit of the controls and the impedances for a single converter connected to an infinite voltage source.}
    \label{fig:SingleConv}
\end{figure}

Transient analysis using $P$–$\delta$ curves relies on an infinite-voltage source to set the reference angle. Fig.~\ref{fig:SingleConv} shows the simplified schematic of a converter connected to a voltage source via a line used for our theoretical analysis. The control system, comprising a voltage loop, a reactive power outer loop, a power synchronization loop, and the VVI, generates the voltage reference for the MMC, represented as a voltage source in series with an impedance. 

To simplify the analysis, the virtual impedance can be added to the electrical circuit. The resulting circuit is shown in Fig.~\ref{fig:SingleConv2} below. $E\angle\delta$ is the voltage reference given by the voltage loop control, and can be found in Fig.~\ref{fig:blckdgrm}.

\begin{figure}[htbp]
    \centering
    \includegraphics[width=1\linewidth]{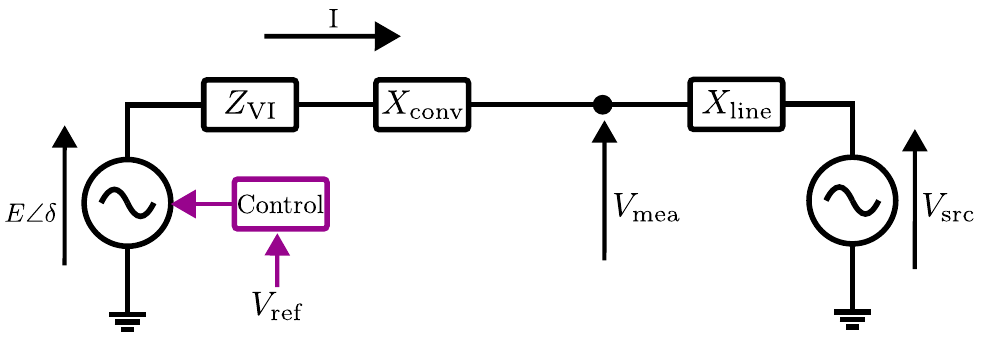}
    \caption{Equivalent circuit of the single converter connected to an infinite bus used for the mathematical analysis. The VVI was added to the electrical circuit, while $V_{\text{valv}}$ is replaced by $E\angle \delta$.}
    \label{fig:SingleConv2}
\end{figure}

Based on the equivalent circuit, the apparent power can be expressed as
\begin{align}
\label{1}
S = V_{\text{src}} I^{*},
\end{align}
 where $()^*$ denotes the complex conjugate, and the current injection is given by
\begin{align}
\label{current}
I = \frac{E\angle \delta - V_{\text{src}}}{R + jX},
\end{align}
where $S \in \mathbb{C}$ and $I \in \mathbb{C}$ are phasor representations of the complex power and the converter current, while $V_{\text{src}} \in \mathbb{R}_{>0}$ is the infinite bus voltage with the reference angle of 0.

Assuming that $R \in \mathbb{R}_{>0}$ and  $X \in \mathbb{R}_{>0}$ represent the total impedance between $E\angle \delta$ and $V_{\text{src}}$ (i.e., the sum of the virtual impedance, the converter's internal impedance and the line impedance), and substituting \eqref{current} into \eqref{1}, we obtain
\begin{align}
S = V_{\text{src}} \frac{E\angle (-\delta) - V_{\text{src}}}{R - jX}.
\end{align}
Using standard properties of complex numbers, this expression becomes
\begin{align}
S = \frac{V_{\text{src}} (R + jX)}{R^{2} + X^{2}}
\left(E\cos\delta - jE\sin\delta - V_{\text{src}}\right).
\end{align}
By taking the real part of this expression, the active power is found to be
\begin{align}
\label{eq:p}
P = \frac{V_{\text{src}} R}{R^{2} + X^{2}}\left(E\cos\delta - V_{\text{src}}\right) + \frac{V_{\text{src}} X}{R^{2} + X^{2}} E\sin\delta .
\end{align}
However, the virtual impedance $Z_{VI}$ is a function of the current magnitude through $R= R(|I|)$ and $X=X(|I|)$. Considering the definition of $\phi(I)$ in \eqref{eq:phi}, the resulting relationships are
\begin{subequations}\label{RX}
\begin{align}
R(|I|) &= \phi(I) R_{\text{VI,init}} + R_{\text{line}} \\
X(|I|) &= \phi_X(I) X_{\text{VI,init}} + X_{\text{line}} + X_{\text{conv}}
\end{align}
\end{subequations}
where the term $\sigma(\phi(I)-1+\frac{1}{\sigma})$ has been grouped into the function $\phi_X(I)$.

\begin{figure}[htbp]
    \centering
    \includegraphics[width=1\linewidth]{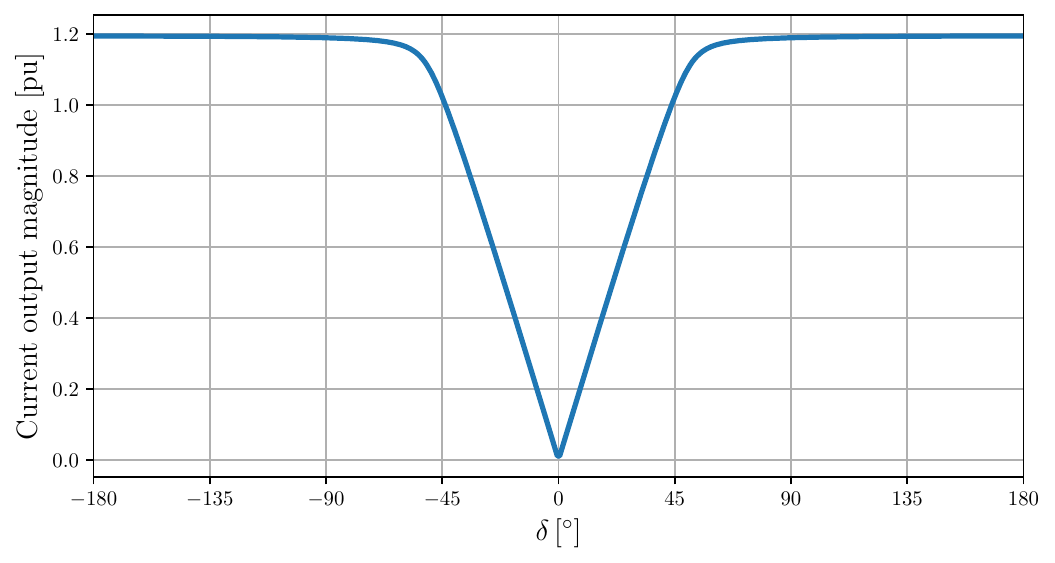}
    \caption{Current magnitude for different power angles $\delta$ under the proposed current limiting control.}
    \label{fig:current}
\end{figure}

Finally, taking the magnitude of \eqref{current}
\begin{align}
\label{eq:I}
|I| = \sqrt{\frac{E^2 + V_{\text{src}}^2 - 2EV_{\text{src}}\cos\delta}{R(|I|)^2 + X(|I|)^2}}
\end{align}
gives the relation between $|I|$ and $\delta$. Plotting the solution of \eqref{eq:I} on Fig.~\ref{fig:current} shows that the current is well limited to the maximum current $I_{\max}=1.2$.

Then, substituting \eqref{eq:I} in \eqref{eq:phi} and \eqref{eq:phi} into \eqref{RX} and substituting those equations in \eqref{eq:p}, we obtain a non-linear relation between $P$ and $\delta$, that can be numerically solved.

\subsection{Transient Stability Analysis of VVI using $P-\delta$ curves}
\begin{figure}
    \centering
    \includegraphics[width=1\linewidth]{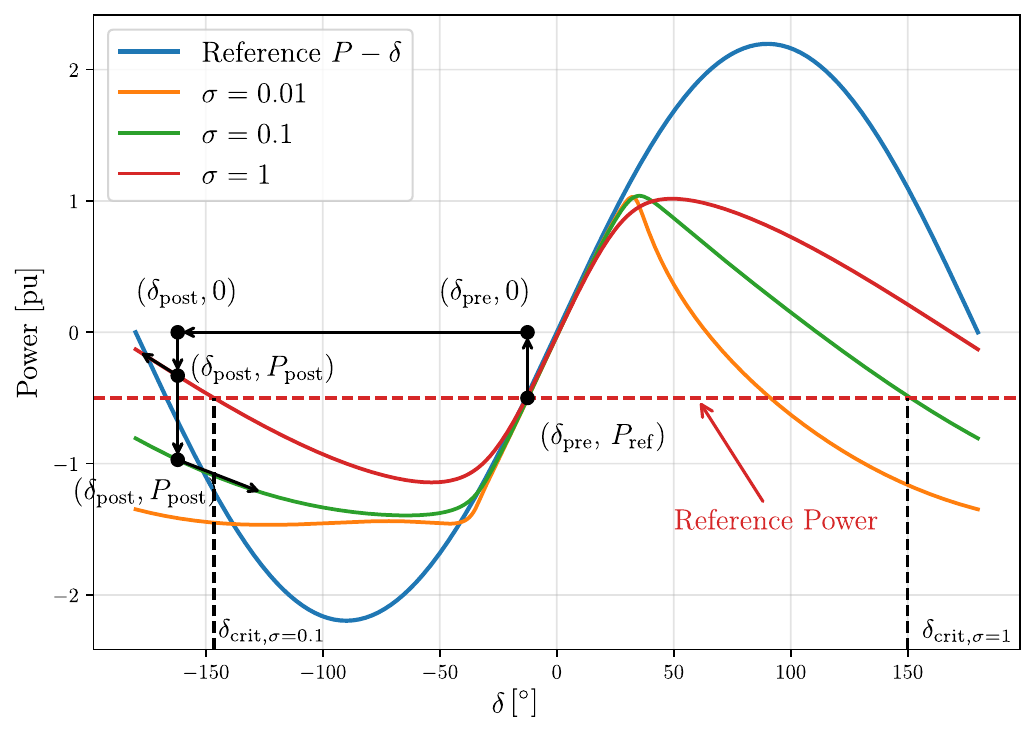}
    \caption{$P-\delta$ Curves with VVI ($\sigma$ sweep) and fault trajectory for a converter absorbing power. Immediately after fault inception, the converter operating point shifts to $(\delta_{\text{pre}}, 0)$ and, during the fault, the angle decreases until the fault is cleared at $(\delta_{\text{post}}, 0)$. If $\delta_{\text{post}}$ exceeds the critical angle  $\delta_{\text{crit}}$, $\delta$ will continue to decrease. If the fault is cleared before $\delta_{\text{post}}<\delta_{\text{crit}}$, then the angle will increase starting from $(\delta_{\text{post}}, P_{\text{post}})$ after the fault is cleared. Notably, a more resistive VVI (i.e., smaller $\sigma$) increases $\delta_{\text{crit}}$.} 
    \label{fig:SigmaSweep}
\end{figure}

Figure~\ref{fig:SigmaSweep} plots the $P-\delta$ curves from \eqref{eq:p}, including a comparison for different values of the increase in the VVI reactance-resistance ratio $\sigma$.
The sign convention for power is positive when the converter injects power into the AC grid (inverter) and negative when it absorbs power (rectifier). When injecting power, a more inductive VVI leads to a larger stability margin, as it moves the $P-\delta$ curve up in the region corresponding to power injection into the AC grid (i.e., $\delta>0$). This increases the so-called decelerating region in the equal-area criterion, as discussed in \cite{Baeckeland2024OvercurrentDiscussion, Ordono2024CurrentThrough, Qoria2023VariableImprovement, Yang2024Small-SignalStabilization}. However, much less attention has been given to the negative part of the $P-\delta$ curve, when the converter is absorbing power from the AC grid (i.e., $\delta<0$).

Since a power synchronization loop without inertia is employed, the classical interpretation in terms of accelerating and decelerating areas is no longer applicable. Instead, the phase angle dynamics are governed by the power error: the angle increases when the measured power is below the reference and decreases when it exceeds the reference. Loss of synchronism occurs if, after fault clearance, the phase angle exceeds the critical angle, defined as the intersection of the $P-\delta$ curve with the reference power. This is illustrated in Fig.~\ref{fig:SigmaSweep} using a short-circuit fault as an example. The subscript "pre" refers to pre-fault conditions, while the suffix "post" refers to post-fault conditions, once the fault is cleared. 

The point ($\delta_{\text{pre}}, P_{\text{ref}}$) represents the steady-state operation. Immediately after a bolted fault occurs, the system moves to point ($\delta_{\text{pre}}, 0$), where power is no longer transferred. Since $P_\text{mea} = 0$ while $P_\text{set} < 0$, the droop law, in red in Fig.~\ref{fig:blckdgrm},
\begin{equation}
    \omega = \omega_\text{ref} + K_{\text{p}}\underbrace{(P_\text{set} - P_\text{mea})}_{\displaystyle <\, 0}
\end{equation}
yields $\omega < \omega_\text{ref}$, causing the virtual angle $\delta$ to decrease until the fault is cleared (point ($\delta_{\text{post}}, 0$)), then returns to its respective $P$--$\delta$ curve, i.e., ($\delta_{\text{post}}, P_{\text{post}}$) corresponding to the VVI incremental reactance-to-resistance ratio $\sigma$.

In the first case, with a more inductive VVI (i.e., larger $\sigma$), the system loses synchronism, and the angle $\delta$ continues to decrease until it reaches the steady-state ($\delta_{\text{pre}}, P_{\text{ref}}$) again. This causes a significant power swing and inversion of power flow, which can damage equipment and/or trip protective relays. On the other hand, using a more resistive VVI (i.e., smaller $\sigma$), the critical angle $\delta_{\text{crit}}$ decreases, resulting in an increased critical clearing time. Notably, for faults cleared before $\delta_{\text{crit}} \leq \delta$ the system returns to normal operation without losing synchronism because $\delta$ is increasing towards ($\delta_{\text{pre}}, P_{\text{ref}}$) after fault clearance.

From this perspective, an optimal $P$–$\delta$ characteristic is one that attains the lowest $P$ in the power-absorbing region (i.e., $\delta<0$), thereby maximizing the allowable angle excursion before reaching the critical angle $\delta_{\text{crit}}$.
Additionally, \cite{Ordono2024CurrentThrough} highlights the damping properties of a resistive VVI as it damps the post-fault oscillations. Notably, making the VVI more inductive can lead to oscillations and harmonic instability~\cite{Yang2024Small-SignalStabilization}.
A fully resistive impedance, however, is not desirable, as power transfer capability is limited at a lower reactance-to-resistance ratio $X/R \in \mathbb{R}_{>0}$, leading to instability for larger resistances and power transfers~\cite{Vilmann2022FrequencyImbalance}. An incremental reactance-to-resistance ratio of $0.1$ is chosen for the remainder of this paper.

\subsection{Transient Stability Analysis using $P-\delta$ curves for VVI combined with VP}
\begin{figure}[htbp]
    \centering
    \includegraphics[width=0.97\linewidth]{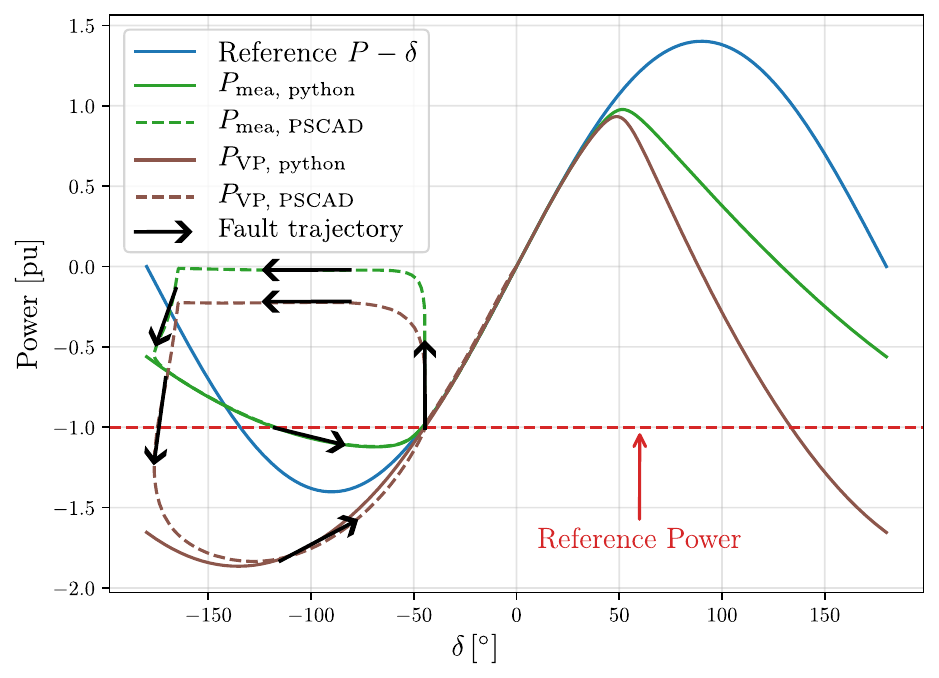}
    \caption{$P-\delta$ curves including VVI and VP using $\sigma = 0.1$. Including the power dissipated in the VVI virtually increases the converter's absorbed power and, hence, the clearing angle. Notably, using only VVI, the critical angle at which $P_{\text{mea}}$ is equal to the reference power is approximately $\delta_{\text{crit},VVI}=-120^\circ$. In contrast, using VVI and VP, the critical angle equals $135^\circ$ (or $-225^\circ$). Fault trajectories obtained using EMT simulations highlight that the angle $\delta$ increases after fault clearing despite having already crossed $\delta_{\text{crit},VVI}$.}
    \label{fig:PDcurveVP}
\end{figure}

As described earlier, the proposed control uses the VVI's virtual power dissipation in the synchronization loop. 
The initial reason for introducing this method is developed in Sec.~\ref{PostFault}, but it appears to offer a significant advantage in terms of the system's transient stability. When this additional element is incorporated into the $P$–$\delta$ characteristic, the resulting curves shown in Fig.~\ref{fig:PDcurveVP} are obtained. Compared to the measured $P-\delta$ curve using VVI (noted $P_{\text{mea}}$) shown in Fig.~\ref{fig:SigmaSweep}, the combined VVI and VP $P-\delta$ curves  (noted $P_{\text{VP}}$) lie significantly below the measured power curve $P_{\text{mea}}$, which implies a larger critical clearing angle and, consequently, improved transient stability. The $P_{\text{VP}}$ curve remains, as expected, close to the $P_{\text{mea}}$ curve during steady-state operation ($|I|<1$), since the extra term in \eqref{VirtualPower} is 0 at low current ($\phi(I)=1$ at low currents).

The dotted lines in Fig.~\ref{fig:PDcurveVP} correspond to PSCAD EMT simulation results using a 78-cell switching MMC model. A three-phase-to-ground fault with a duration of 0.25 s is applied to the grid ($V_{\text{src}} = 0$ on Fig.~\ref{fig:SingleConv}). After fault clearance, the system successfully maintains synchronism due to the inclusion of the virtual power term. In the absence of this mechanism, the system would experience a power inversion, as the measured power $P_{\text{mea}}$ exceeds the reference power after fault clearance, ultimately leading to loss of synchronism.

After fault clearance, the small deviation between the EMT simulation results (dashed lines) and the theoretical predictions (solid lines) arises from the voltage controller and the low-pass filter on the power measurement, which are neglected in the mathematical formulation. Their presence introduces a time delay, explaining the difference between the simulation and the theoretical results.

\begin{figure}[htbp]
    \centering
    \includegraphics[width=1\linewidth]{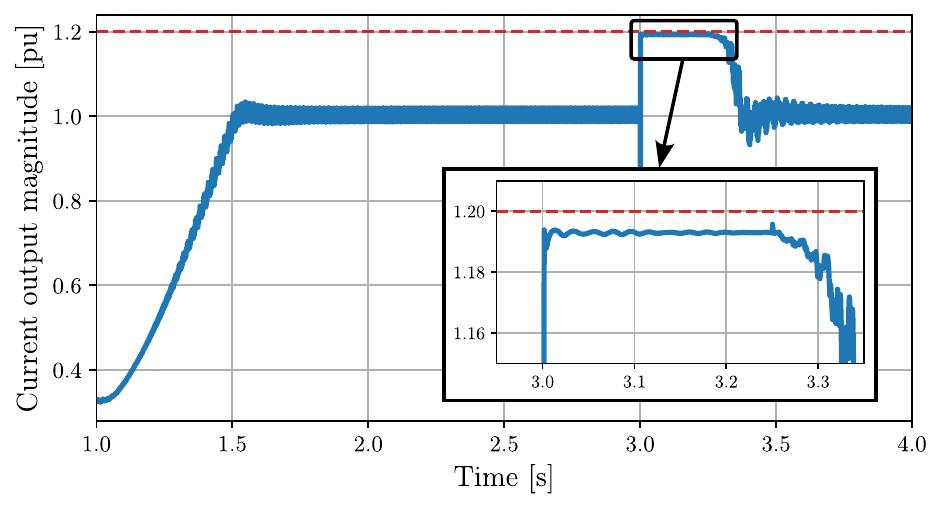}
    \caption{Response of a 78-cell switching MMC model to a symmetric ground fault at $t=3$~s. The current magnitude is limited to $I_{\max}=1.2$~pu and returns to the pre-fault current magnitude after the fault is cleared at $t=3.25$~s.}
    \label{fig:currentTime}
\end{figure}

Fig.~\ref{fig:currentTime} presents the converter's current from the same EMT simulation during the fault. It shows that the method effectively limits the current to below the set limit (1.2 pu in this case). 

\subsection{Post-Fault scenario}\label{PostFault}
In this post-fault scenario, conventional VVI control is no longer suitable. After synchronization is re-established, power transfer among converters is governed exclusively by the power synchronization mechanism rather than by the electrical impedances between units. 
As a solution, the virtual losses in the virtual resistance are added to the measured power, increasing the power perceived by the synchronization loop. Adding the new power equation \eqref{VirtualPower} to the droop equations \eqref{eq:omega} results in
\begin{align}
\label{eq:droopinit}
    \omega_i = \omega_0 + m_i (P_{i,\mathrm{ref}} - P_i + (\phi(I_i) - 1)R_{\text{VI,init}}|I_i|^2)
\end{align}
for all $i \in \{1,\ldots,n\}$. Assuming $1$~pu voltage and only d-axis current, the current equals the power in pu, hence $\phi(I)=\phi(P)$.
In the steady state, we obtain 
\begin{align}
\label{eq:droop equations}
    \omega_i = \omega_0 + m_i (P_{i,\mathrm{ref}} - P_i + (\phi(P_i) - 1)R_{\text{VI,init}}P_i^2)
\end{align}
for all $i \in \{1,\ldots,n\}$ as well as 
\begin{subequations}
\label{eq:PowerAndFreq}
    \begin{align}
    \omega_1 = \omega_2 = \dots = \omega_n,\\
    \sum\nolimits_{j=1,n} P_j = P_{\text{WF}},
\end{align}
\end{subequations}
i.e., frequency synchronization and power balance equations. The system of equations formed by \eqref{eq:droop equations} and \eqref{eq:PowerAndFreq} can be solved and plotted for different wind farm powers, $P_{\text{WF}}$, to analyze the post-fault steady-state power sharing, as illustrated in the case study of Sec.~\ref{sec:CaseStudy}.

\section{Case study: Offshore Energy Hub}
\label{sec:CaseStudy}
To illustrate the effectiveness of this method, a $4$~GW offshore energy hub is used. Four $1$~GW wind farms (aggregated model) are connected to four offshore converters, divided into two $2$~GW bi-poles. The converters are used to export offshore power to the onshore grid via HVDC lines, as shown in Fig.~\ref{fig:OEH} presented in the introduction.
The wind farms are modeled as power sources, including $P$ and $Q$ outer loops and an inner current loop. They synchronize to the grid via a PLLs.
The detailed parameters used in the simulation are listed in Table~\ref{tab:system_parameters}.

\subsection{Simulation results}
\begin{figure}
    \centering
    \includegraphics[width=1\linewidth]{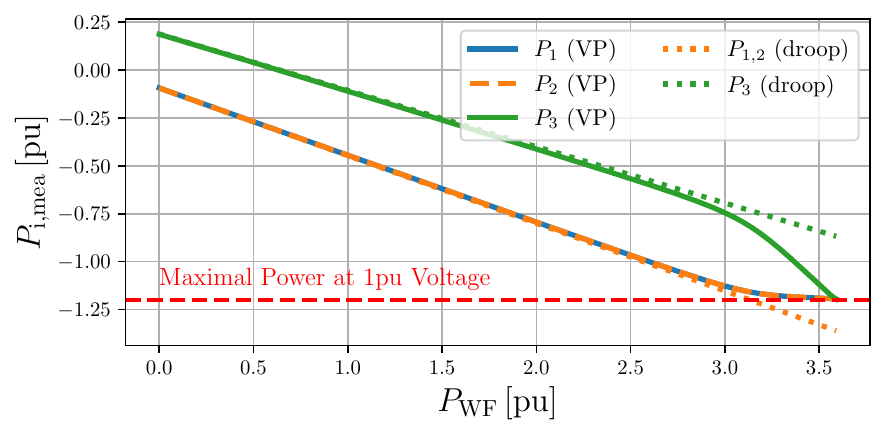}
    \caption{Power sharing in a three-converter system. Assuming $1$~pu voltage, the power is similar to the current (in pu). Beyond $P_{\text{WF}} \approx 3$, overload of converter \#1 and \#2 is mitigated by increasing the power absorbed by converter \#3 relative to droop-based power sharing.}
    \label{fig:frequency}
\end{figure}

The scenario includes different initial power setpoints and droop coefficients, listed in Table~\ref{tab:conv_params}. After the system reaches steady state, converter \#4 is disconnected from the grid by opening the transformer's circuit breaker (in green in Fig.~\ref{fig:OEH}), and its pre-fault power must be shared among the three remaining converters. The solution of~\eqref{eq:droop equations} is presented in Fig.~\ref{fig:frequency} for $n=3$, representing the post-fault scenario.

\begin{table}[htbp]
\centering
\caption{Droop settings for the HVDC converters}
\label{tab:conv_params}
\footnotesize
\setlength{\tabcolsep}{7pt}
\renewcommand{\arraystretch}{1.1}
\begin{tabular}{@{}c c c @{}}
\toprule
 & \textbf{Conv 1 \& 2} & \textbf{Conv 3 \& 4} \\
\midrule
\textbf{$P_{i,\mathrm{ref}}$} & -0.8 & -0.4 \\
\textbf{$m_i$}                & 0.05 & 0.06 \\
\bottomrule
\end{tabular}
\end{table}

The plot shows that no converter exceeds the maximum power (i.e., current under the $1$~pu voltage assumption) and is automatically dispatched across the different connected converters, unlike when droop control is used. 
However, it is assumed that, after fault clearance, the total power supplied by the wind farms remains within the aggregate power capacity of the OEH after losing any single converter (i.e., $N-1$ criterion). This assumption is reasonable, as exceeding this limit would inevitably lead to converter overloading regardless of the control strategy, effectively driving the VVI to extreme values for all units. In such a scenario, additional power curtailment or dissipation mechanisms would be required, either on the offshore AC side or at the wind farm level.

A time-domain simulation is conducted to illustrate the benefits of the proposed method. At $t=3.5$~s, converter \#4 is disconnected from the grid using the green circuit breaker (CB$_4$) in Fig.~\ref{fig:OEH}, creating a sudden power imbalance and exposing the remaining converters to potential overcurrent conditions.

\begin{figure}[htbp]
    \centering
    \begin{subfigure}[t]{0.9\linewidth}
        \centering
        \includegraphics[width=\linewidth]{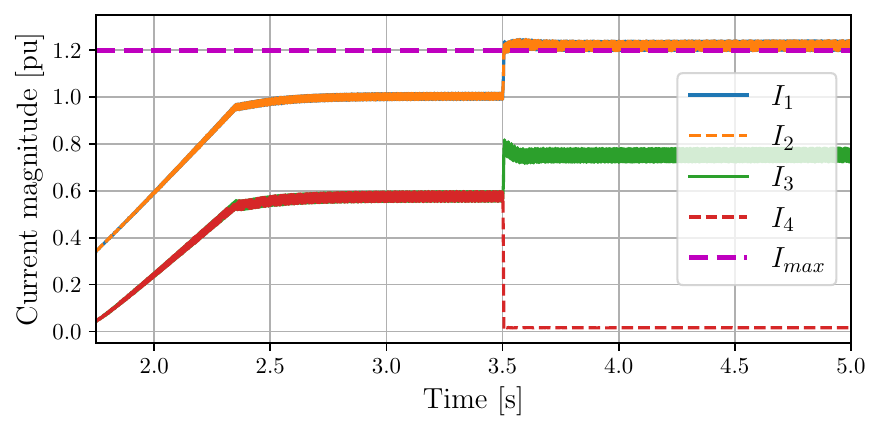}
        \caption{Only droop control (no VVI or VP).         \label{fig:curr:droop}}
    \end{subfigure}
    
    \begin{subfigure}[t]{0.9\linewidth}
        \centering
        \includegraphics[width=\linewidth]{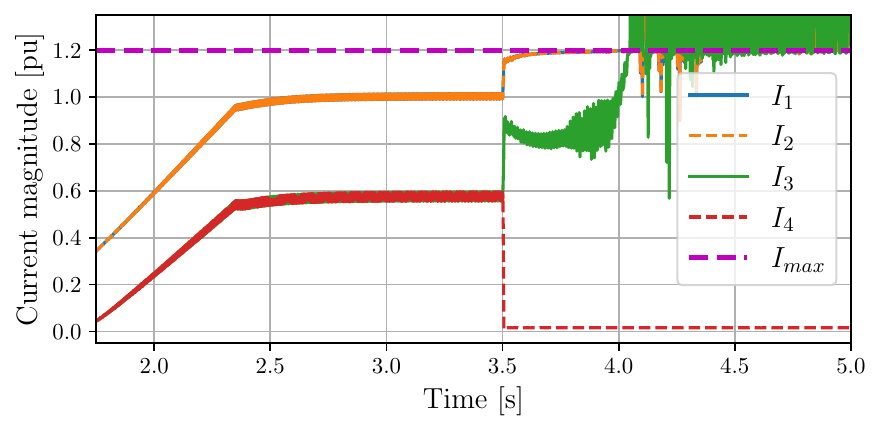}
        \caption{Droop with VVI (no VP).         \label{fig:curr:novp}}
    \end{subfigure}

    \begin{subfigure}[t]{0.9\linewidth}
        \centering
        \includegraphics[width=\linewidth]{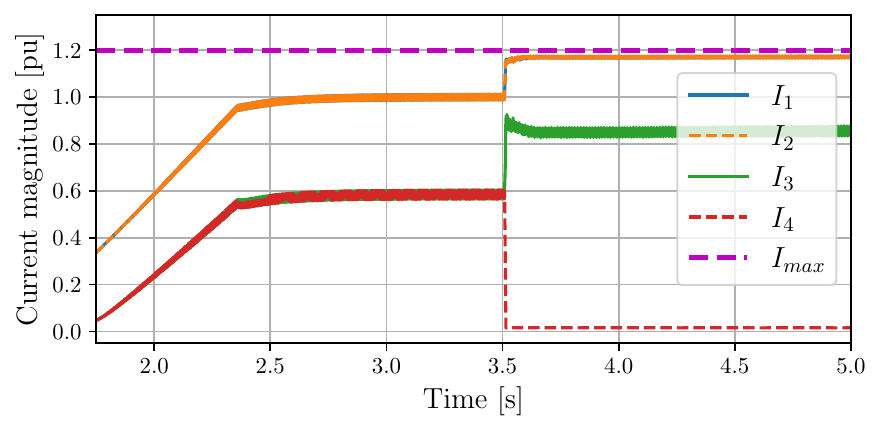}
        \caption{Droop with VVI and VP.         \label{fig:curr:vp}
}
    \end{subfigure}

    \caption{Converter current after disconnecting converter \#4 with different control strategies. When only grid-forming droop control (a) is used, converters \#1 \& \#2 exceed their current limit. If droop control with only VVI is used (b), the VVI increases too much, leading to system collapse. Finally, if droop control with VVI and VP (c) is used, none of the converters exceeds its current limit.}
    \label{fig:curr:all}
\end{figure}

Without any current-limiting or overload-mitigation strategy, i.e., when using a conventional droop-based control as presented in \cite{Ghimire2023Grid-formingWPPs,Guerrero2011HierarchicalStandardization}, converters \#1 and \#2 exceed their current limits. In this case, power sharing is governed solely by the initial droop settings (see Fig.~\ref{fig:curr:droop}). This case is included solely for illustrative purposes. We highlight that, even though the current limit violation is small in this outage scenario, this control cannot be used in practice because it would cause damage to the semiconductor switches during more severe contingencies (e.g., short-circuit faults). 

If only the VVI were applied, the system would become unstable due to an increasing VVI value, which would counter the synchronization loop and force a defined power transfer. This is shown in Fig.~\ref{fig:curr:novp}. In contrast, incorporating the VP term into the synchronization loop enables a coordinated redistribution of power: converter \#3, which initially operates with a significant margin, automatically increases its power contribution. This results in a stable operating point in which the load is shared according to the remaining converters' available capacity. Simulations results for the proposed control including both VP and VVI, are shown in Fig.~\ref{fig:curr:vp}.

\section{Conclusion}
This paper presented a unified grid-forming control framework that improves both transient stability and autonomous post‑fault power redistribution in fully converter‑based grids typically encountered in offshore energy hubs and offshore wind farms. The proposed variable virtual impedance, implemented via a smooth threshold function, limits current while preserving the converter’s voltage‑source behavior during severe disturbances, such as short-circuit faults and a loss of a converter station. In contrast to conclusions typically drawn for converters injecting AC power, our analysis demonstrated that a more resistive virtual impedance provides superior transient stability compared to inductive virtual impedance when converters absorb AC power (i.e., operate as rectifiers). This behavior was analytically explained in terms of the negative‑power region of the $P-\delta$ characteristics.

As the VVI cannot effectively mitigate overcurrents during overpower events, a novel virtual-power mechanism, computed from the dissipated virtual-resistance power, is introduced. This term effectively serves as an autonomous, coordinated post‑fault power‑sharing mechanism, allowing the remaining converters to autonomously absorb excess power without modifying droop gains or set‑points. At the same time, the virtual power term enhances transient stability during short-circuit faults by increasing the perceived synchronizing power in the grid-forming controller when the converter approaches its current limit. 

The combined VVI–VP strategy was validated through EMT simulations in PSCAD, which confirmed its ability to (i) tightly limit fault currents, (ii) maintain synchronism during severe three‑phase faults, and (iii) achieve proportional and stable power redistribution after converter disconnection. Overall, the results demonstrate that the proposed method offers a robust, scalable, and analytically tractable solution to current limitations and to stability enhancement in emerging grids dominated by power electronic converters.


\section*{Declaration of generative AI and AI-assisted technologies in the manuscript preparation process}
During the preparation of this work, the author(s) used Deepl Write and Grammarly in order to improve clarity and style. Also, ChatGPT and Copilot were used to speed up coding. Perplexity was used to support and find relevant papers for the literature review. After using these tools/services, the author reviewed and edited the content as needed and takes full responsibility for the content of the published article.


\section*{Appendix}
\begin{table}[htbp]
    \centering
    \caption{System parameters for the offshore wind farm HVDC transmission system used for the study case in Sec.~\ref{sec:CaseStudy}.}
    \label{tab:system_parameters}
    \begin{tabular}{|l|l|r l|}
        \hline
        \multicolumn{4}{|c|}{\textbf{Wind Farms}} \\
        \hline
        \multicolumn{2}{|l|}{Voltage} & 66   & kV                  \\ \hline
        \multicolumn{2}{|l|}{Base Power}   & 1000 & MW                  \\ \hline
        \multicolumn{2}{|l|}{$X$}     & 0.15 & pu                  \\ \hline
        \multicolumn{2}{|l|}{$R$}     & $1\times10^{-4}$ & pu      \\
        \hline
        \multicolumn{4}{c}{}                                        \\[-1pt]
        \hline
        \multicolumn{4}{|c|}{\textbf{Offshore AC Grid}} \\
        \hline
        \multicolumn{2}{|l|}{Voltage} & 66   & kV                  \\ \hline
        \multirow{3}{*}{$\pi$-line} & $L$   & 0.2  & mH    \\ \cline{2-4}
                                            & $R$   & 0.02 & $\Omega$ \\ \cline{2-4}
                                            & $C/2$ & 23   & $\mu$F   \\
        \hline
        \multicolumn{4}{c}{}                                        \\[-1pt]
        \hline
        \multicolumn{4}{|c|}{\textbf{HVDC Converters}} \\
        \hline
        \multicolumn{2}{|l|}{Voltage}              & 300  & kV          \\ \hline
        \multicolumn{2}{|l|}{Base Power}              & 1000 & MW           \\ \hline
        \multicolumn{2}{|l|}{$X_{\text{conv}}$}    & 0.05 & pu          \\ \hline
        \multicolumn{2}{|l|}{$K_\text{p}$}         & 1    & pu          \\ \hline
        \multicolumn{2}{|l|}{$K_\text{i}$}         & 50   & s$^{-1}$    \\ \hline
        \multicolumn{2}{|l|}{$X_{\text{VI,init}}$} & 0.4  & pu          \\ \hline
        \multicolumn{2}{|l|}{$R_{\text{VI,init}}$} & 0.04 & pu          \\ \hline
        \multicolumn{2}{|l|}{$K_q$}                & 0.02 & pu          \\ \hline
        \multicolumn{2}{|l|}{$\sigma$}             & 0.1  & --          \\ \hline
        \multirow{2}{*}{$\phi(|I|)$} & $K_p$          & 10  & --   \\ \cline{2-4}
                                     & $I_\text{max}$ & 1.2 & pu   \\
        \hline
        \multicolumn{4}{c}{}                                        \\[-1pt]
        \hline
        \multicolumn{4}{|c|}{\textbf{Transformers}} \\
        \hline
        \multicolumn{2}{|l|}{Base Power}              & 1000 & MW           \\ \hline

        \multirow{2}{*}{66/400 kV}  & $X$ & 0.1   & pu \\ \cline{2-4}
                                    & $R$ & 0.01  & pu \\ \hline
        \multirow{2}{*}{400/300 kV} & $X$ & 0.15  & pu \\ \cline{2-4}
                                    & $R$ & 0.015 & pu \\
        \hline
    \end{tabular}
\end{table}


\bibliographystyle{IEEEtran}
\bibliography{references.bib}
\end{document}